\documentclass[twocolumn,showpacs,preprintnumbers,amsmath,amssymb]{revtex4}

\begin{document}

\title{Collapsing sphere on the brane radiates}

\author{M Govender$^a$}
\email{megang@dit.ac.za}
\author{N Dadhich$^{b,c}$}
\email{nkd@iucaa.ernet.in}
\affiliation{$^1$Department of Physics, Durban Institue of Technology, P O Box 953, Durban, South Africa}\
\affiliation{$^2$Inter-University Centre for Astronomy and Astrophysics, Post Bag 4, Ganeshkhind, Pune-411 007, India}\
\affiliation{$^3$School of Mathematical and Statistical Sciences, University of Natal, Durban, 4041, South Africa}

\begin{abstract}
We study the analogue of the Oppenheimer-Snyder model of a collapsing sphere
of homogeneous dust on the Randall-Sundrum type brane. We show that the
collapsing sphere has the Vaidya radiation envelope which is followed by
the brane analogue of the Schwarzschild solution described by the
Reissner-Nordstr${\ddot o}$m metric. The collapsing solution is matched to 
the brane generalized Vaidya solution and which in turn is matched to the 
Reissner-Nordstr${\ddot o}$m metric. The mediation by the Vaidya radiation 
zone is 
the new feature introduced by the brane. Since the radiating mediation is 
essential, we are led to the remarkable conclusion that a collapsing 
sphere on the brane does indeed, in contrast to general relativity, 
radiate null radiation.
\end{abstract}

\pacs{04.50.+h; 04.70.Bw; 98.80.Cq; 12.10.-g; 11.25.Mj}

\maketitle


The seat of high energy modifications to general relativity (GR) would be
the neighbourhoods of singularities occurring in gravitational collapse and
the cosmological big-bang explosion. The study of situations leading to such
singular events would therefore be most pertinent in the models that claim
to incorporate such modifications. The Randall-Sundrum (RS) brane world 
model \cite{r1} is one such model which has currently attracted great 
attention and 
activity. In this paper, we wish to study
gravitational collapse of a homogeneous dust sphere on the RS type brane. 
In GR, it is described by the well-known
Oppenheimer-Snyder (OS) model \cite{n1} in which the interior of 
collapsing sphere
is given by the Friedmann metric while the exterior is the static 
Schwarzschild solution. The two are matched continuously on the moving
boundary. This reflects the well-known classical result that a 
collapsing sphere does not radiate. In contrast it turns out that a 
collapsing dust sphere on the brane would require in general a non static 
exterior \cite{r}, and that could be the Vaidya radiating solution on the 
brane \cite{n2}.
That is, a collapsing sphere must have a radiation envelope 
described by the brane analogue of the Vaidya radiating solution \cite{n2}, 
which could finally be matched to the brane analogue of the Schwarzschild 
solution described by the Reissner-Nordstr${\ddot o}$m (RN) metric 
\cite{r6}. We shall thus establish that the analogue of the 
Oppenheimer-Snyder collapse on the brane is a Vaidya radiating sphere.

String theory and M-theory represent one of the routes to find a
covering theory for GR, which would possibly be a theory of quantum gravity.
In this approach, gravitation becomes a higher
dimensional interaction with the 4-D conventional GR as its low energy
limit. In the brane world scenario, the matter fields remain confined to the
3-brane, which is our 4-D Universe we live in, while gravity can propagate
in higher dimensional spacetime called the bulk. In the RS brane model,
gravity is localized on the brane by the curvature of
the bulk spacetime which is an Einstein space with negative $\Lambda$.
What essentially happens is that gravitational field gets "reflected" back
onto the brane through the Weyl curvature of the bulk as a trace free matter
field. It is termed the Weyl (dark) radiation on the brane. This is the
non-local effect mediated by the bulk Weyl
curvature while the local effect manifests through the extrinsic curvature
of the brane resulting into square of the stresses.
This leads to an effective Einstein 
equation on the brane \cite{r2} which
incorporates both local as well as non-local effects representing the high
energy modifications of GR. At the very basic level the brane modification to 
the Newtonian potential goes as $r^{-3}$, which is symptomatic of the anti 
de-Sitter (AdS$_{5}$) bulk.

A solution of the effective vacuum equation on the brane for a static black
hole was obtained \cite{r6}, and it turned out to be the RN metric of a charged
black hole in GR. Here charge refers not to electric charge but instead to
the Weyl tidal charge as a measure of the reflected gravitational field
energy from the bulk. (We would refer it as the Weyl RN (WRN) solution.) 
Finding the corresponding solution in the bulk and then to
match it
onto the black hole solution on the brane with proper boundary conditions
is a very difficult and formidable task. The evolution and extension 
of the brane solution into the bulk numerically as well as analytically is 
also wrought
with considerable difficulties (\cite{r8} and \cite{rsd}). As a first step, 
the consideration
is therefore focussed on solving the effective equation on the brane without
reference to the bulk spacetime except that it is taken to have non-zero 
Weyl curvature.

The question of gravitational collapse of a homogeneous dust sphere on the 
brane has very recently been addressed \cite{r} in which it was concluded 
that the exterior spacetime has to be non-static unless the collapsing sphere is 
of pure Weyl 
radiation. A model of a star in hydrostatic equilibrium on the brane has also 
been considered \cite{r7}. 
In that the interior is the brane analogue of the Schwarzschild interior 
solution of uniform density while the exterior is no longer unique as was the 
case in GR. This is because the exterior spacetime is now not vacuum. In the 
context of the cosmic censorship conjecture, collapse of the Vaidya null 
radiation onto an empty cavity on the brane has also been studied \cite{n2}. 
It is 
shown that the brane effects favour formation of a black hole against a naked 
singularity. This is in line with the expectation that the brane effects tend 
to strengthen the gravitational field. In all these studies, only the 
effective equation on the brane has been solved without reference to the bulk 
spacetime except that it is assumed to have  
non-zero Weyl curvature to project dark radiation on the brane. We would also 
resort to the same strategy and would only address the effective Einstein 
equation on the brane.

Since the exterior has to be non-static as well as free of collapsing matter, 
it could only have radial energy flux which could sustain the non-static 
character. The two possiblities are the Vaidya null radiation and the heat 
flux. The latter is ruled out because there cannot exist a spherically 
symmetric solution with heat flux alone as the source. This leaves 
only the null radiation for the exterior, which is described by the Vaidya solution 
on the brane \cite{n2}. 
In this paper we wish to show that the brane analogue of the 
Oppenheimer-Snyder (OS) collapse of homogeneous dust sphere is: Friedmann 
collapsing metric for the interior which is matched to the Vaidya radiating envelope,
which in turn matches finally to the asymptotically flat WRN metric. 
Here and henceforth all solutions would refer to their brane analogues, while 
the analogue of the Schwarzschild would be referred explicitly as WRN. This is the 
canonical paradigmical picture of gravitational collapse on the brane.

The effective Einstein field equation on the brane is given by \cite{r2}
\begin{equation} 
G_{\mu\nu} = -\Lambda g_{\mu\nu} + 8\pi GT_{\mu\nu} + \frac{48\pi G}{\lambda}S_{\mu\nu} - {\cal E}_{\mu\nu}\label{01}\end{equation}
where $\lambda > 10^8 GeV^4$ is the brane tension, $\Lambda$ is the brane
cosmological constant, $S_{\mu\nu}$ represents the quadratic stresses and
$\cal E_{\mu\nu}$ is the projection of the Weyl curvature on the brane
giving the Weyl radiation. The Bianchi identities imply the following
conservation laws:
\begin{equation}\nabla^{\nu}T_{\mu\nu} = 0, \hspace{5mm} \nabla^{\nu}{\cal E}_{\mu\nu} = \frac{48\pi G}{\lambda}\nabla^{\nu}S_{\mu\nu}
, \hspace{5mm} {\cal E}_{\mu}{}^{\mu} = 0\end{equation}
The above equations do not however close because of the presence of 5-D
degrees of freedom in $\cal E_{\mu\nu}$. For this, the Lie derivatives of
the brane should be considered \cite{r2}, which would then close the system. We
shall however not address the full 5-D equations and would rather
confine ourselves to the brane degrees of freedom.


Now we consider a collapsing dust cloud described by the Friedmann metric in 
isotropic coordinates
\begin{equation} \label{1}
ds^2 = -d\tau^2 + \frac{a(\tau)^2}{(1 + \mbox{$\frac{1}{4}$}kr^2)^2}\left[dr^2 + r^2 d\Omega^2\right]\end{equation}
where $\tau$ is the proper time. 
The proper radius is given by 
\[R(\tau) = \frac{ra(\tau)}{1 + \mbox{$\frac{1}{4}$}kr^2}\]. 
The modified Friedman equation can be written as
\begin{equation} \label{2}
\frac{{\dot a}^2}{a^2} = \frac{8\pi G}{3}\rho\left[1 + \frac{\rho}{2\lambda}\right] + \frac{C}{\lambda a^4} - \frac{k}{a^2} + \frac{\Lambda}{3}\end{equation}
where $C$ is the Weyl radiation constant fixed by the bulk Weyl curvature, and the ${\rho}^2$ term is the local high energy correction. The ususal Friedmann 
collapse is recovered when $\lambda^{-1}\rightarrow 0$. From the conservation
equation (2), it follows that $\rho= \rho_0(a_0/a)^3$ where $a_0$ is the 
intial radius from which the collapse began.

Since the collapsing boundary surface $\Sigma$ is free falling we can write
\[R_{\Sigma}(\tau) = \frac{r_0a(\tau)}{(1 + \mbox{$\frac{1}{4}$}kr_0^2)^2}\]
where $r = r_0 = {\mbox constant}$. This allows  us to recast (\ref{2}) into  
\begin{equation} \label{222}
{\dot R}^2 = \frac{2GM}{R} + \frac{3GM^2}{4\pi \lambda R^4} + \frac{Q}{\lambda R^2} + E + \frac{\Lambda}{3}R^2 . \end{equation}
In the above equation the total energy per proper stellar volume, $M$ and the total "tidal charge" $Q$ are given by \cite{r}
\begin{equation}
M = \frac{4\pi a_0^3 r_0^3 \rho_0}{3(1 + \mbox{$\frac{1}{4}$}kr_0^2)^3}, \hspace{2cm} Q = C \frac{r_0^4}{(1 + \mbox{$\frac{1}{4}$}kr_0^2)^4}\end{equation}
and the "energy" per unit mass assumes the following form 
\begin{equation}
E = -\frac{kr_0^2}{(1 + \mbox{$\frac{1}{4}$}kr_0^2)^2}, \hspace{2cm} E > -1 .
\end{equation}
The brane contribution to the collapse comes through $\rho^2$ and the 
Weyl radiation. It would turn out that it is the latter which is 
responsible for the non-static exterior. As argued above, the only possible 
choice for the exterior is the Vaidya null radiating spacetime. On the 
brane it would be obtained by solving the equation,
\begin{equation}
G_{\mu\nu} = \Lambda g_{\mu\nu} -8\pi\sigma k_\mu k_\nu + {\cal E}_{\mu\nu}
\end{equation}
where $k_\mu k^\mu = 0$ which implies $S_{\mu\nu} = 0$. 
For the exterior, we seek the general solution (for the Weyl stresses, as for 
WRN, the null energy condition, ${\cal E}_{\mu\nu} k^\mu k^\nu = 0$ is 
assumed) of this equation which is given by \cite{n2},
\begin{eqnarray} \label{3}
ds^2 &=& - \left(1 - \frac{2Gm(v)}{\sf r} - \frac{Q(v)}{{\sf r}^2} - \frac{\Lambda {\sf }^2}{3}\right)dv^2 \nonumber \\
&&+ 2dvd{\sf r} + {\sf r}^2d\Omega^2\end{eqnarray}
where $v$ is the retarded Eddington null coordinate. We would employ the 
standard 4-D Israel matching conditions on the brane which 
would require continuity of the metric and the extrinsic curvature of the 
boundary surface $\Sigma$. That would mean continuity of the metric 
components and of $\dot R$. Apart from these conditions, we also have one more 
condition coming from vanishing of fluid pressure, $p=0$.

In matching of metrics (\ref{1}) and (\ref{3}) we employ the procedure due to 
Santos \cite{san}.
For the metric (\ref{3}) the equation of the surface $\Sigma$ is given by
\[f({\sf r}, v) = {\sf r} - {\sf r}_{\Sigma}(v) = 0\].
The first junction condition requires that the metric functions match smoothly across the boundary surface $\Sigma$. 
For the metrics (\ref{1}) and (\ref{3}) this yields
\begin{eqnarray*} 
d\tau &=& \left(1  -\frac{2Gm(v)}{{\sf r}_{\Sigma}} - \frac{\Lambda{\sf r}_{\Sigma}^2}{3} - \frac{Q(v)}{{\sf r}_{\Sigma}^2} + 2\frac{d{\sf r}}{dv}\right)^{\frac{1}{2}} dv \label{333a}\\ \nonumber \\
R_{\Sigma} &=& {\sf r}_{\Sigma}(v). \label{333b}
\end{eqnarray*}
The second matching condition requires that the extrinsic curvature components $K_{ij}$ are continuous across $\Sigma$ which would be equivalent to continuity 
of $\dot R$. For the metric (9), we write 
\[{\dot R}^2 = \frac{2Gm(v)}{R} + E + \frac{\Lambda}{3}R^2 + \frac{Q(v)}{\lambda R^2} \]
on $\Sigma$, 
which when compared to (5) yields the condition
\begin{equation}
m(v) = M + \frac{3M^2}{8\pi \lambda R^3} + \frac{1}{2RG\lambda}\left(Q - Q(v)\right) \label{555a} \end{equation}
where we have also used the continuity of the metric. Eqn. (5) is the same as 
eqn. (7) of \cite{r} while eqn. (10) is the analogue of eqn. (15) of \cite{r}.

Apart from this we also have fluid pressure $p = 0$, which would further 
yield the condition
\begin{equation} Q(v) = \left[\frac{3GM^2}{4\pi R^2} + Q \right]_{\Sigma} \label{555b}\end{equation}
Now eqn. (10) simply gives $m(v) = M$, which is the analogue of the 
OS condition $m(R) = M$ in GR. The new condition which marks the brane effects is the above equation (11). It is this condition which implies that a collapsing sphere on the brane 
must radiate because $Q(v)$ represents propagation of nonlocal stresses. On 
the other hand $Q(v) = 0$ would imply both $Q = 0$ (if projection of the bulk 
Weyl vanishes in the exterior so should for the interior as well) and $M = 0$,
as predicted in \cite{r}. There it was obtained by requiring $R = 4\Lambda$ 
in the exterior. This is the main result which simply follows from 
the matching conditions and the pressure free condition. The result would 
be true even when pressure is non-zero because on the boundary it will always 
vanish. A collapsing fluid sphere on the brane would therefore radiate and its
exterior will have the radiation envelope described by the Vaidya 
solution.

Finally we shall match the Vaidya solution to the WRN static metric which 
could be done in a very straight forward manner. 
We now match the metric (9) to the WRN metric given by
\begin{equation} \label{6}
ds^2 = - A dt^2 + \frac{d r^2}{A} + r^2d\Omega^2 \end{equation}
where
\[A = 1 - 2\frac{G{\cal M}}{r} - \frac{q}{r^2} - \frac{\Lambda}{3}{r}^2\].
The junction conditions required for the smooth matching of 
metrics (9) and (12) and their associated extrinsic curvature components yield
\begin{eqnarray}
m(v) &=& {\cal M} + \frac{q}{2Gr} - \frac{Q(v)}{2Gr} \label{7a}\\ \nonumber \\
m(v) &=& {\cal M} + \frac{q}{Gr} - \frac{Q(v)}{Gr} .\label{7b}
\end{eqnarray}
Comparing (\ref{7a}) and (\ref{7b}) we obtain the expected result 
\begin{equation}
m(v) = {\cal M},  ~ Q(v) = {q}\end{equation}

This completes the matching exercise. It shows that a collapsing Friedmann 
sphere (OS collapse) on the brane has Vaidya radiation envelope followed by 
the WRN metric.


In the recent study of OS-like collapse on the RS type brane \cite{r}, it
was concluded that exterior to the collapsing sphere cannot be static
unless the collapse is that of pure Weyl radiation. We have argued that the  
non static character of exterior spacetime could only be sustained by 
the Vaidya null radiation flowing out radially from the collapsing sphere. 
We have therefore taken for exterior the radiating generalization of the 
static black hole solution on the brane as given in \cite{n2}. This is at one 
end matched to the Friedmann solution for interior of the collapsing sphere 
and at the other to the WRN static metric for the black hole on the brane 
\cite{r6}. The matching conditions relate the mass functions at the two 
boundary surfaces, and the new brane effects are negotiated by the Weyl 
radiation and the density square contributions. It is interesting to note 
that for a collapsing sphere to be non-vacuous, what is required is the 
Vaidya null radiation along with the Weyl radiation contribution. That is, 
the bulk must have non-zero Weyl curvature. The distinguishing feature of 
spherical collapse on the brane is that 
{\em it radiates out null radiation and requires the bulk spacetime 
to be conformally non-flat}. 

From physical considerations as alluded earlier, it is clear that the Vaidya 
radiating solution is perhaps the correct and natural choice for the exterior 
of the collapsing sphere on the brane. Since it is simply the radiating 
generalization of WRN, it would also not have the right weak field limit of 
the brane modification to the Newtonian potential going as $r^{-3}$. The cause
 of the problem lies in the non-zero Weyl curvature of the bulk, which gives 
rise to the Weyl tidal charge producing $r^{-2}$ term in the potential. In 
the Vaidya solution, null radiation part which is brane bound and hence is 
correctly incorporated. The problem is with the Weyl radiation stresses 
which are projected from the bulk and hence cannot truly be specified without 
reference to the bulk spacetime. This is the crux of the problem which runs 
through all the solutions on the brane. Note that in the Friedmann solution 
too, the equation of state for radiation is imposed on the Weyl stresses. 
They are, without reference to bulk spacetime, only supposed to be trace-free.
 The question would not therefore be resolved without the knowledge of bulk 
spacetime. For gaining some insight into the bulk, there have been 
attempts to analytically and numerically Taylor expand the brane solutions 
into the bulk \cite{n8,n9,rsd}. In the absence of full knowledge of the bulk 
spacetime, this is the only way open to get some useful knowledge of the bulk 
in the close vicinity of the brane corresponding to the brane solutions.  

Further it may also be noted that the RS brane world model is very critically 
tuned to conformally flat ADS$_5$ bulk \cite{sd,n9}. For instance, it has 
been shown that for a conformally non-flat bulk, there exists no bound state 
for zero mass gravitons on the brane \cite{sd}. This is the key feature of 
the RS brane model which seem to be so tightly tied to asymptotically AdS bulk \cite{pn}. 
Also recall that the weak field limit is obtained for a Minkowski brane 
with conformally flat ADS$_5$ bulk. It would therefore be fair to say that 
the question in its entirity is perhaps open. This is of course an issue of 
viewpoint which would have various shades of belief and perception.

The Weyl tidal charge in WRN is essentially the measure of gravitational
field energy being "reflected" from the bulk on the brane as a trace-free
matter field. It is expected to be negative \cite{r2,r6,d}. It turns out that 
the brane modifications to GR would strengthen gravity (i.e. contribute in 
line with GR) only when tidal charge is negative \cite{d}. This is the most 
interesting aspect of the brane world formulation in which bulk contributes 
actively to the dynamics of the brane. For this bulk must have non-zero Weyl 
curvature. Also note that in the Vaidya generalization of WRN, not only the 
mass $M$ but the tidal charge $Q$ also becomes function of retarded null 
coordinate $v$. This happens because $Q$ is essentially caused by $M$ (through 
the reflected field energy from the bulk) and hence it would be related to it 
\cite{r2,r6,d}.

The canonical picture that emerges for the spherical collapse on the brane 
is that the collapsing sphere has the radially propagating null radiation
envelope followed by the static spacetime of a black hole on the brane. This 
is in contrast to the classical picture where a spherically collapsing or 
oscillating system is not supposed to radiate out energy. This happens 
because of the bulk gravitons projected stresses on the brane and also of  
effective pressure not being zero on the boundary. That means dynamically the 
system is, in contrast to the classical case, not closed and consequently 
physical information could travel from interior to exterior. This is why the 
exterior cannot be static. Though we have only considered homogeneous dust 
collapse, this feature would be true in general for collapse on the brane.
Since it is radiative for spherical case, it would certainly be so for more 
general collapse. This is the general feature of collapse on the brane. 

We have however not addressed the 5-D equations for determination of  
the bulk spacetime. When the problem is fully solved by incorporating the 
bulk spacetime, the canonical picture as alluded above would hold good 
qualitatively. What would change would be the exact form of the solutions like 
WRN and its Vaidya generalization. The picture, the collapsing sphere having 
a null radiation envelope followed by static spacetime of a black hole on the 
brane, would stay. It is remarkable that without reference to bulk spacetime, 
we are able to deduce qualtatively the distinguishing characteristic features 
of gravitational collapse on the brane.

\begin{acknowledgments}

The major part of this work was done when ND was visiting the 
University of Natal, Durban. It was sparked off by ND's 
discussions with Roy Maartens at GR-16.
We thank Roy for setting us on the problem through exciting 
discussions as well as insightful critical comments on the manuscript and 
also Kesh Govinder for useful discussions. ND would like to thank Sunil 
Maharaj and the Relativity group at Durban for warm hospitality which has 
facilitated this work.

\end{acknowledgments}

\end{document}